# Objective Measures of Perceptual Audio Quality Reviewed: An Evaluation of Their Application Domain Dependence

Matteo Torcoli , Thorsten Kastner , and Jürgen Herre , *Senior Member, IEEE*

*Abstract*—Over the past few decades, computational methods have been developed to estimate perceptual audio quality. These methods, also referred to as objective quality measures, are usually developed and intended for a specific application domain. Because of their convenience, they are often used outside their original intended domain, even if it is unclear whether they provide reliable quality estimates in this case. This work studies the correlation of well-known state-of-the-art objective measures with human perceptual scores in two different domains: audio coding and source separation. The following objective measures are considered: fwSNRseg, dLLR, PESQ, PEAQ, POLQA, PEMO-Q, ViSQOLAudio, (SI-)BSSEval, PEASS, LKR-PI, 2f-model, and HAAQI. Additionally, a novel measure (SI-SA2f) is presented, based on the 2f-model and a BSSEval-based signal decomposition. We use perceptual scores from 7 listening tests about audio coding and 7 listening tests about source separation as ground-truth data for the correlation analysis. The results show that one method (2f-model) performs significantly better than the others on both domains and indicate that the dataset for training the method and a robust underlying auditory model are crucial factors towards a universal, domain-independent objective measure.

## I. INTRODUCTION

**B**ASIC Audio Quality (BAQ) defines a general, domain-independent quality criterion to rate the perceived overall quality of a signal being tested [1]. BAQ is one of the main evaluation criteria in audio coding and was used also as assessment criterion in the field of blind source separation [2]. Listening tests (often referred to as subjective evaluation, e.g. MUSHRA [1]) in a controlled environment are the most reliable method for assessing BAQ. These are, however, time-consuming and costly and cannot be easily carried out at each development stage, e.g. of a new audio codec. The recent social distancing measures due to the COVID-19 pandemic have added another difficulty to conducting listening tests in the laboratory. Therefore, objective evaluation measures are greatly desired, i.e. computational methods that are able to estimate BAQ as closely as possible to the human assessment [3].

These models are usually designed and trained on audio material and distortion types representing the specific domain in which the measures are intended to be used. A good measure is expected to generalize to audio material unseen during its development, as long as the distortions typical of the application domain are encountered. A universal measure would generalize also to unseen distortions from different application domains. Universality or domain-independence is often implied to a certain degree, even without clear evidence that this is a valid assumption. An interesting example is represented by PESQ [4], which was finalized 20 years ago for the evaluation of speech quality over telephony systems. It is nowadays widely used for evaluating methods potentially introducing very different types of distortion, e.g. speech separation based on Deep Neural Networks (DNNs), singing voice extraction, and dereverberation, e.g. for hearing aids [5]–[9]. PESQ has also been proposed as loss function for supervised learning [10], [11].

The correlation between objective measures and perceptual scores has been studied by many authors, but usually within a specific domain application or with limited amount of perceptual ground-truth data [7], [12]–[27].

The aim of this paper is to shed some light on these issues with the following contributions:
- State-of-the-art intrusive objective measures are reviewed, with a glance at DNN-based non-intrusive estimates.
- The correlation with ground-truth data from 7 listening tests about audio coding and 7 listening tests about source separation is analyzed. The prediction generalization on different domains is investigated. The used listening tests are based on MUSHRA, which is suited for assessing intermediate quality of audio signals. A generalization of the results on estimating BAQ of signals with small impairments should not be done without further research.

- A novel measure based on the 2f-model, preceded by a BSSEval-based signal decomposition is presented for assessing the perceptual relevance of artifacts.

## II. OBJECTIVE MEASURES

This section reviews state-of-the-art objective measures. The focus of this work is on intrusive measures, i.e. ideal reference signals are used for comparison to estimate the audio quality of the signal under test. The only novel contribution of this section is given in Section II-M. A reader with little time and previous knowledge of the state of the art can skip the rest and continue with Section III.

The measures described from Section II-A to Section II-C belong to the speech enhancement domain, while the ones from Section II-D to Section II-G were developed in the field of audio coding. The measures from Section II-H to Section II-M focus on source separation. Section II-N introduces HAAQI, which was designed for hearing aids. Finally, Section II-O discusses the recent developments leveraging deep learning.

Many of the following measures originally support only mono signals. In this case, we compute the mean of the per-channel outputs when dealing with stereo signals.

### A. Frequency-Weighted Segmental SNR (fwSNRseg)

The fwSNRseg [28] quantifies the power ratio of the reference signal and a noise signal that is obtained as the difference of the reference and the test signal. The fwSNRseg is computed and weighted for each time frame and each subband of a filterbank with critical-band spacing. The implementation in [5] is used, where the weights are computed from the subband-magnitude of the reference raised to the power of 0.2.

### B. Log-Likelihood Ratio Distance (dLLR)

The dLLR [29] is based on the assumption that, over short time intervals, speech can be represented by an all-pole model. Linear Prediction Coefficients (LPC) are computed for the test signal and the reference. The two LPC sets predict the reference with certain residual energies. The dLLR is defined as the logarithm of the ratio of these residual energies. We employ the implementation in [5], where the distance is limited to 2 before averaging over time.

### C. Perceptual Evaluation of Speech Quality (PESQ)

PESQ [4], [30], [31] was designed for speech transmitted over telecommunication networks and narrow-band speech codecs. The method comprises a pre-processing that mimics a telephone handset. Measures for audible disturbances are computed from the specific loudness of the signals and combined in PESQ scores. From these, a Mean Opinion Score (MOS) [32] is predicted by means of a polynomial mapping function. We use the wideband mode of the ITU reference software [4]. This operates at a sampling frequency of 16 kHz. So the signals are resampled to this sampling frequency before they are fed to the tool. Stereo signals are supported natively. The tool exited with a *processing error* for 8% of the signals in the PEASS datasets (details in Section III-B). These signals are discarded in the following correlation analysis of PESQ.

### D. PEAQ (Perceptual Evaluation of Audio Quality)

PEAQ [33], [34] is a measurement scheme for the perceptual evaluation of coded audio signals. Several mid-level perceptual features, called Model Output Variables (MOVs), are derived by either comparing the error signal with estimated masking thresholds or by comparing the internal ear representations of reference and test signal. They are combined in a neural network computing the main output, i.e. the Overall Difference Grade (ODG). Two versions of PEAQ are defined: 1) the Basic version, designed for applications requiring high processing speed, and 2) the Advanced version, designed for higher accuracy at the expense of speed. We use the Basic version by the McGill University, publicly available as MATLAB code [35]. Multi-channel signals are natively supported. The individual MOVs were also shown to be good predictor of perceived audio quality for different tasks [21], [36]. We consider the MOVs that exhibited the highest correlation performance in our experiments as well as in [21]: Average Distorted Blocks (ADB), Noise-to-Mask Ratio (NMR), Windowed average of the Modulation Difference #1 (WinModDiff1), and Average Modulation Difference #1 (AvgModDiff1).

### E. Perceptual Objective Listening Quality Assessment (POLQA)

POLQA [23], [37] was developed as a *"technology update"* for PESQ and it was designed to predict the perceived overall speech quality of listening tests that comply with [32] or [38] (the test signals used in this work do not necessarily meet this requirement). POLQA operates in two modes: narrowband or superwideband. We use a proprietary implementation licensed by OPTICOM in the superwideband mode and compare the three main versions of POLQA: Version 1.1 (01/2011), Version 2.4 (09/2014), and Version 3 (03/2018) [39].

### F. Perception Model-Based Quality (PEMO-Q)

PEMO-Q [40] aims to be a general measure of perceived audio quality for any type of audio signals and audio signal distortions. It is an extension of a previous work on speech quality assessment [41]. The measurement scheme compares the internal ear representations of the reference and the test signal like PEAQ and POLQA. The internal representations are estimated using a psychoacoustic model [42]. Three-dimensional (time, frequency, and modulation) internal representations of the signals are obtained and the cross-correlation coefficient between the test and reference representations is calculated and used as a measure of the perceived similarity, i.e. the Perceptual Similarity Measure (PSM). A regression function based on subjective data is then applied to map the PSM to the ODG. For consistency with PEASS (Section II-J), we use the PEMO-Q version used by PEASS and publicly available from [43].

### G. ViSQOLAudio

ViSQOLAudio [44] is a metric designed for estimating the quality of general coded audio at 48 kHz developed from Virtual Speech Quality Objective Listener (ViSQOL) [45], [46], which was focused on speech signals. Both metrics are based on a model of the peripheral auditory system to create spectro-temporal internal representations of the signals called neurograms. These are compared via an adaptation of the structural similarity index, originally developed for evaluating the quality of compressed images and then adapted to predict intelligibility [47]. Version 3 was recently released [48], [49] and it is here referred to as ViSQOLAudioV3. The declared aim for this new version is to *"fill the blind spots in the training/validation datasets"* so as to have a more general system that would perform better *"in the wild"*. This tool internally down-mixes multi-channel signals to mono.

### H. Blind Source Separation Evaluation (BSSEval)

BSSEval [50] is a multi-criteria performance evaluation toolbox. The toolbox is widely used in the source separation community and it was used as main figure of merit in several community-based evaluation campaigns from 2007 [51] to 2018 [52]. BSSEval projects the estimated source onto the subspace spanned by all reference source signals, including filtered versions of those. This filter can be time-variant and its length can be adjusted by the user. In practice, a time-invariant 512-tap-FIR-filter is normally used. The estimated signal is thereby decomposed into target signal $s_{\text{target}}$ and components, meant to be related to different types of error: spatial distortion ($e_{\text{spatial}}$), interference from other sources ($e_{\text{interf}}$) and projection error, interpreted as artifacts ($e_{\text{artif}}$). Energy-based signal-to-error ratios are computed from these components and expressed in dB. Two modes are available: sources (only for mono sources) and source images (i.e. multi-channel sources). We use the images mode, but we do not consider the spatial distortion, i.e. $s'_{\text{target}} = s_{\text{target}} + e_{\text{spatial}}$. Source to Distortion Ratio (SDR), Source to Interference Ratio (SIR) and Source to Artifact Ratio (SAR) are considered. We use version 3.0 of the Matlab toolbox [53]. We limit the range of the output metrics to $[-30 \text{ dB}, 30 \text{ dB}]$.

### I. Scale-Invariant (SI) BSSEval

Starting from the premise that BSSEval has *"generally been improperly used and abused, resulting in misleading results"*, modified and simpler definitions for the BSSEval measures were proposed in [54]. These are called scale-invariant (SI), i.e. SI-SDR, SI-SAR, SI-SIR, and they are particularly recommended by their authors for single-channel separation evaluation. The main difference to BSSEval is the usage of a single coefficient $\alpha$ to account for scaling discrepancies instead of the full 512-tap filter. Hence, a broadband scaling is the only forgiven difference with the reference signal. The measures are defined in a way for which the following relationship holds: $10^{-\text{SI-SDR}/10} = 10^{-\text{SI-SIR}/10} + 10^{-\text{SI-SAR}/10}$. We use our own implementation of the measures, done following the description in [54]. For stereo signals (not covered in the original paper) we compute $\alpha$ considering all channels during the projection, as per BSSEval. Also in this case, we limited the range of the output metrics to $[-30 \text{ dB}, 30 \text{ dB}]$.

### J. Perceptual Evaluation Methods for Audio Source Separation (PEASS)

PEASS [55] was proposed for the perceptual assessment of separated audio source signals, developed as a perceptually motivated successor of BSSEval. Perceptual similarity scores are computed in a two-stage fashion: Error signals reflecting different types of signal distortion are computed from the estimated source signal similar to BSSEval but in a time-frequency selective manner using a gammatone filterbank. Differently from BSSEval, the perceptual salience of the error signals reflecting target source, interfering source, and artifacts is assessed with PEMO-Q, i.e. considering the perceptually relevance of the error signals. For this purpose, a reference signal is generated first for each error type by subtracting the according error signal from the estimated source signal. The resulting perceptual similarity scores ($q^{\text{overall}}$, $q^{\text{target}}$, $q^{\text{interf}}$, $q^{\text{artif}}$) are mapped using a small neural network trained with subjective ratings to form measures of different perceptual audio quality. Herein, Overall-, Artifact- and Interference-related Perceptual Score (OPS, APS and IPS) are considered. We use Version 2.0.1 of the PEASS software [43], where a 2-layer neural network is used for generating the output metrics. This version includes substantial differences with respect to the original proposal in [55]. The authors report that these modifications *"greatly improve correlation with human assessments"*. Multi-channel signals are natively supported.

### K. Log Kurtosis Ratio Perceptually Improved (LKR-PI)

LKR-PI [56] is a measure of perceived musical artifacts (a.k.a. artificial noise, musical noise, or birdies) caused by spectral holes or islands. It is based on the measured change in spectral kurtosis between before and after processing. The change in spectral kurtosis is measured in a black-box fashion, i.e. without assumptions on the distribution of the signal power spectra. This is combined with a perceptually motivated pre-processing. For the source separation domain, the measurement is performed only on the leaking interferer, where the spectral kurtosis should not change. This excludes the portions of the signal where the target source is active. In the correlation analysis of LKR-PI, we ignore the signals for which the excluded portion is bigger than 95% of the total length. About 40% of the signals in the following evaluation have to be discarded for this domain, while no signal is excluded for the audio coding domain.

### L. The 2f-Model

The 2f-model [16] estimates the perceived quality of separated source signals, driven by 2 MOVs from PEAQ Basic: ADB and AvgModDiff1. ADB estimates the amount of noticeable distortions in units of the just noticeable level difference between test and reference signal. AvgModDiff1 assesses differences in the temporal modulation of the loudness envelopes between the signals. The two MOVs are computed with the PEAQ Basic version provided by the McGill University [35], which is publicly available. The parameters for combining these MOVs and so obtaining the final 2f-model score were newly computed for this implementation and are available online [57]. This differs slightly from the original proposal in [16], in which an internal PEAQ implementation (and so a different set of combining paramers) was used.

### M. SI-SA2f

For BAQ any perceived deviation from the reference signal is considered a degradation. However, it is often of interest to assess the presence of artifacts independently from the interferer reduction, e.g. in applications such as source separation. For this purpose, we propose to combine the signal decomposition used in SI-BSSEval and the perceptual model offered by the 2f-model. We name this novel measure SI-SA2f (i.e. a blend of SI-SAR and 2f-model) and it has the same aim as (SI-)SAR, APS, and LKR-PI, i.e. assessing the amount of artifacts independently of the interferer reduction. Starting from the signal under test $y$ and the ground-truth sources, the SI-BSSEval signal decomposition provides the following signals: $s_\text{target}$, $e_\text{interf}$, $e_\text{artif}$, where $y = s_\text{target} + e_\text{interf} + e_\text{artif}$. SI-SA2f is obtained by running the 2f-model on the signal under test $y$ and using $s_\text{target} + e_\text{interf}$ as reference signal. Hence:

- If $e_\text{artif} \to 0$ then SI-SA2f$\to 100$.
- If $e_\text{artif} \to y$ (and $s_\text{target} \to 0$) then SI-SA2f$\to 0$.

### N. Hearing-Aid Audio Quality Index (HAAQI)

HAAQI [58] was designed to predict music quality for individuals listening through hearing aids. The index is based on a model of the auditory periphery [59], extended to potentially include the effects of hearing loss. This is fitted to a dataset of quality ratings made by listeners having normal or impaired hearing. The rated signals feature musical content, modified by different types of processing found in hearing aids. Some of these processes are common also in audio coding and source separation. The hearing loss simulation can be bypassed and the index becomes valid also for normal-hearing people; we use HAAQI in this normal hearing. An implementation provided by the original author is used in this investigation. Based on the same auditory model, the authors of HAAQI also proposed a speech quality index (HASQI) and a speech intelligibility index (HASPI): references are given in [58].

### O. Methods Based on Deep Learning

All measures reviewed so far originate outside the deep learning paradigm. DNNs have gained a lot of momentum over the past few years and DNNs for estimating the perceived audio quality were proposed. Two main approaches can be identified. The first one uses a large amount of subjective perceptual scores to train new DNN-based objective measures [60]. The second approach is to train DNNs to estimate existing measures (such as the ones described in the previous sections), e.g. with the aim of making them completely or partially non-intrusive [61]–[64]. Of the referenced works, only [63] provides the trained DNNs [65], referred to as Waveform Evaluation Networks (WEnets). These are four DNNs, trained to predict PESQ, POLQA, PEMO-Q, or Short-Time Objective Intelligibility (STOI), without reference signals. The four networks were here tested (with input level normalization active, 3-seconds segments, 50% overlap, and averaging over the estimated scores for one items). The DNN achieving the best overall performance is reported in the following. This is the one predicting PESQ and is referred to as *WEnets PESQ*.

## III. Ground-Truth Subjective Scores

This section describes the datasets of subjective reference ratings, which will be used as ground truth for the correlation analysis in the following sections. An overview of these datasets is given in Table I together with the number of ratings from each listening test. We consider 7 listening tests in the audio coding domain (from 2 independent sources) and 7 listening tests in the source separation domain (also from 2 independent sources). The ground-truth perceptual scores consist of averages over all ratings given to each signal.

All the listening tests followed MUSHRA or MUSHRA-like procedures for assessing intermediate quality of audio signals. The perceptual scores of the considered listening tests span the full quality scale, from poor to excellent quality, which is an important factor to consider while interpreting the correlation results in the following sections. Further research is required for domains where only a small portion of the quality scale is spanned or where only small impairments are observed.

### A. Audio Coding

*1) Coding Artifacts [66]:* In this set of listening tests, 16 subjects assessed the quality of signals that were distorted in a controlled fashion with different monaural coding artifacts so to simulate sub-optimal audio coding operating points. Each distortion was applied on a different set of 8 musical signals, with no overlap between sets. The following 5 distortions were considered, each applied with 5 different coarse quality levels:

- Birdies, i.e. warbling artifacts generated by spectral holes or islands.
- Bandwidth limitation (BW Lim), i.e. low-pass-filtered versions with an adapted crossover frequency.
- Pre-echoes, i.e. fuzzy onsets, imprecise percussion timing, and ghost voice for speech signals.
- Tonality or harmonicity mismatch, i.e. simulating a sub-optimal bandwidth extension, where all spectral content above a given crossover frequency was replaced by a scaled copy of the remaining lower part of the spectrum.
- Unmasked noise, i.e. simulating a suboptimal bandwidth extension, where all spectral content above a given crossover frequency is substituted by random noise with the same spectral envelope.

*2) MPEG USAC Verification Test [67]:* Three verification tests were run to assess BAQ of the Unified Speech and Audio Coding (USAC) [67], where USAC was compared with AMR-WB+ and HE-AAC v2 at different bit-rates. We consider Test 1 (USAC t1) and Test 3 (USAC t3). Excluding the items used during the listener training, USAC t1 and USAC t3 contain the same 24 audio excerpts. USAC t1 considers only the first channel of each items and the mono signals are encoded at low bit rates (8-24 kbps). USAC t3 considers the stereo signals and these are encoded at high bit-rates (32-96 kbps). The 24 items comprise music-only samples, speech-only samples, and mixed speech and music samples. Of these, 5 items have no spoken or singing voice at all, while 7 have spoken speech only. The low-pass anchor and the reference conditions are not considered, leaving 9 conditions for USAC t1 and 8 conditions for USAC t3, all being real-world coding conditions. USAC t1 involved 62 listeners from 13 different test sites with previous training and post-screening (but not all of them rated all signals). USAC t3 involved 29 listeners from 6 sites.

TABLE I
OVERVIEW OF LISTENING TESTS USED FOR THE GROUND-TRUTH PERCEPTUAL SCORES. ONLY EXPERT LISTENERS TOOK PART IN THE LISTENING TESTS (EXCEPT SISEC18 DATASET INCLUDING ALSO NON-EXPERT LISTENERS). AMOUNT OF RATED SIGNALS AND SYSTEMS ARE GIVEN WITHOUT CONSIDERING ANCHORS AND REFERENCES. SIGNALS INCLUDE MIXTURES OF SPEECH AND NOISE OR MUSIC, MUSIC WITH SINGING VOICE, OR MUSIC ONLY. THE PERCENTAGE OF ITEMS WHERE THE TARGET CONTAINS ONLY SPOKEN SPEECH IS GIVEN

| Domain | Audio Coding | | | | | | | Source Separation | | | | | | |
|---|---|---|---|---|---|---|---|---|---|---|---|---|---|---|
| Dataset | Coding Artifacts | | | | | MPEG USAC | | PEASS | | SEBASS | | | | |
| Publicly available | No | | | | | Members | | Yes | | Yes | | | | |
| Listening test | Birdies | Tonality Mismatch | BW Lim | Unmasked Noise | Pre-echoes | t1 (mono, low rates) | t3 (stereo, high rates) | OPS LT | APS LT | SASSEC | SiSEC08 | SAOC DB | PEASS BAQ | SiSEC18 |
| Number of systems | 5 | 5 | 5 | 5 | 5 | 9 | 8 | 13 | 13 | 11 | 9 | 9 | 13 | 334 |
| Overall rated signals | 40 (=8x5) | 40 | 40 | 40 | 40 | 219 (=24x9) | 192 (=24x8) | 40 | 40 | 154 | 126 | 126 | 40 | 144 |
| Mono % / Stereo % | 0 / 100 | 25 / 75 | 0 / 100 | 25 / 75 | 75 / 25 | 100 / 0 | 0 / 100 | 20 / 80 | 20 / 80 | 0 / 100 | 0 / 100 | 0 / 100 | 20 / 80 | 0 / 100 |
| Target: spoken speech only % / other % | 0 / 100 | 0 / 100 | 0 / 100 | 0 / 100 | 0 / 100 | 29 / 71 | 29 / 71 | 50 / 50 | 50 / 50 | 57 / 63 | 57 / 63 | 57 / 63 | 50 / 50 | 0 / 100 |
| Ratings per signal | 16 | 16 | 16 | 16 | 16 | 56-58 | 29 | 20 | 20 | 15 | 14 | 6-12 | 7 | 10-19 |
| Total ratings | 640 | 640 | 640 | 640 | 640 | 12,312 | 5,568 | 800 | 800 | 2,310 | 1,764 | 1,146 | 280 | 2,022 |

### B. Source Separation

*1) PEASS [55]:* The PEASS dataset was used for the development of PEASS. The dataset contains separated sources and specifically defined anchor signals including listener ratings on *global quality* (i.e. BAQ), *preservation of the target source*, *suppression of other sources* and *absence of additional artificial noise* for each audio signal. The following evaluation considers the ratings regarding the global quality (referred to as PEASS OPS LT) and the ratings on the absence of additional artificial noise (referred to as PEASS APS LT).

*2) Subjective Evaluation of Blind Audio Source Separation (SEBASS) [16], [57]:* The SEBASS dataset is a collection of five listening tests on BAQ of separated audio sources from blind and informed source separation systems. These listening tests are referred to as: SASSEC, SiSEC08, PEASS BAQ, SiSEC18, and SAOC DB. In each listening test, except SAOC DB, the listeners rated separated signals submitted as part of community-based signal separation evaluation campaigns, as indicated by the names of the datasets. PEASS BAQ contains the signals from the PEASS OPS LT but ratings from [16]. The main difference with PEASS OPS LT in terms of listening test design is that the listeners of PEASS BAQ were not instructed to rate the worst item with 0. Instructing to rate the worst item with 0 is not compliant with MUSHRA. SAOC DB contains scores investigating the influence on quality of a separated source when an enhanced t/f rendering architecture, as it is offered by MPEG Spatial Audio Object Coding (SAOC) [68], is used for acoustic reproduction [69]. Separated source signals from SASSEC were used to drive the enhanced rendering architecture and the resulting signals were evaluated alongside the original separated signals. The ratings relative to the original separated signals are not considered as part of the SAOC DB in the following, as ratings for the same signals are already contained in SASSEC. As a technology, SAOC is an interesting case where (informed) source separation and audio coding overlap [68], [70], [71].

## IV. CORRELATION ANALYSIS: CRITERIA

In order to assess the performance of the considered objective measures (Section II), a correlation analysis of the measures outputs with the subjective scores (Section III) is carried out. For each listening test, Pearson's and Kendall's correlation coefficients are computed.

All signals from the datasets are re-sampled to 48 kHz or to 16 kHz for PESQ and WEnets PESQ (highest supported sampling frequency).

### A. Pearson's and Kendall's Correlations

Given the ground-truth perceptual scores $X$ for a set of signals and the outputs $Y$ from a measure run on the same signals, the Pearson's correlation coefficient $\rho$ is computed as:

$$\rho(X,Y) = \frac{\sum_{i=1}^{N}(X_i - \overline{X})(Y_i - \overline{Y})}{\sqrt{\sum_{i=1}^{N}(X_i - \overline{X})^2}\sqrt{\sum_{i=1}^{N}(Y_i - \overline{Y})^2}}, \quad (1)$$

where $i$ serves as index for the signals in the considered listening test and the over-line indicates the mean over all signals, e.g. $\overline{X} = \frac{1}{N}\sum_{i=1}^{N}X_i$. Pearson's correlation measures linear correlation between $X$ and $Y$, i.e. how close their relationship is to a first-oder polynomial: $\rho(X,Y) = 1$ indicates total positive linear correlation, while $\rho(X,Y) = 0$ indicates no correlation at all and $\rho(X,Y) = -1$ indicates total negative correlation. As we are not interested in distinguishing between positive or negative correlation, but we are interested in how strong the correlation is, the absolute value of $\rho$ is reported, ranging between 0 and 1. Pearson's $\rho$ can be significantly smaller than 1 even with identical ranking between the elements in $X$ and the one of the elements in $Y$. For this reason, we consider also Kendall's rank correlation $\tau$, which is a measure of the ordinal association (or ranking):

$$\tau(X,Y) = \frac{2\,K}{n(n-1)}, \quad (2)$$

TABLE II
AGGREGATED CORRELATION $\overline{\rho}$ AND $\overline{\tau}'$ ON AUDIO CODING AND SOURCE SEPARATION DATASETS (BAQ) FOR SOME SELECTED MEASURES. AGGREGATED SCORES > 0.80 ARE SHOWN IN BOLD

| $\overline{\rho} / \overline{\tau}'$ | Audio coding | Source separation |
|---|---|---|
| 2f-model | **0.90 / 0.91** | **0.86 / 0.82** |
| ADB | **0.86 / 0.88** | **0.83 / 0.81** |
| HAAQI | **0.84 / 0.84** | 0.69 / 0.61 |
| PEAQ ODG | **0.81 / 0.84** | 0.63 / 0.66 |
| PESQ | 0.75 / **0.84** | 0.76 / 0.70 |
| POLQAv3 | 0.60 / 070 | 0.74 / 0.70 |
| PEMO-Q ODG | 0.72 / **0.82** | 0.77 / 0.78 |
| OPS | 0.71/ 0.74 | 0.67 / 0.63 |
| POLQAv1 | 0.69 / 0.72 | 0.74 / 0.68 |
| fwSNRseg | 0.67 / 0.71 | 0.69 / 0.70 |
| POLQAv2 | 0.56 / 0.57 | 0.74 / 0.71 |
| SDR | 0.58 / 0.67 | 0.63 / 0.68 |

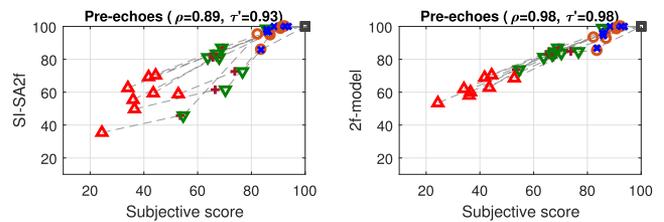

Fig. 1. Difference between SI-SA2f (left) and 2f-model (right) in predicting BAQ for the pre-echoes test. Different symbols are used for the 5 different quality levels. Dashed lines connect the points corresponding to the same items in the different quality levels. The reference signals depicted with black squares are not considered in the calculation of the correlation coefficients.

where K correponds to the number of concordant pairs minus the number of discordand pairs, i.e.:

$$K = \sum_{i=1}^{N-1} \sum_{j=i+1}^{N} c(X, Y, i, j), \quad (3)$$

where $i$ and $j$ serve as indices for the signals in the listening test and $c(X, Y, i, j)$ measures the pairs concordance:

$$c(X, Y, i, j) = \begin{cases} 1, & \text{if } (X_i - X_j)(Y_i - Y_j) > 0 \\ 0, & \text{if } (X_i - X_j)(Y_i - Y_j) = 0 \\ -1, & \text{if } (X_i - X_j)(Y_i - Y_j) < 0 \end{cases} \quad (4)$$

In order to make the values of $\tau$ more comparable with the ones of $\rho$, $\tau$ is mapped to $\tau'$ and the absolute value of $\tau'$ is reported in the following. The mapping is as follows [72]:

$$\tau' = \sin\left(\tau \frac{\pi}{2}\right). \quad (5)$$

These two metrics ($\rho$, $\tau'$) have the advantage of being scale-independent, which is desired in our analysis, in which measures with outputs on different scales are compared. E.g. PEASS and the 2f-model range from 0 to 100 (MUSHRA scores), while PEAQ estimates an ODG ranging from $-4.0$ to $0$. Other metrics (such as the ones suggested in [73], e.g. the Root Mean Square Error) are not scale-independent and are not adopted here.

The statistical significance for the correlation coefficients is also tested (t-test, two-tailed, $\alpha = 0.05$). Tables III, IV, and V report an asterisk on the coefficients for which the null hypothesis could be rejected.

### B. Aggregated Scores

A meta-analysis is conducted where the correlation coefficients for a number of experiments (i.e. a subjective data pool) are aggregated in one score, referred to as *aggregated score*. This aggregation is done by applying the Fisher-z transform on the correlation coefficients, calculating the mean in this domain (where the sampling distribution of the resulting coefficients is approximately normal), and inverting the transformation [74].

The coefficients of to the datasets used during the development of a given measure are not considered in the computation of the aggregated score for that measure. This is noted by (†) next to the ignored coefficients in Tables III, IV, and V. The Fisher-z transform is defined as:

$$z = \frac{1}{2} \ln\left(\frac{1+\gamma}{1-\gamma}\right), \quad (6)$$

where $\gamma$ can be either $\rho$ or $\tau'$. If $\gamma = \rho$, the aggregated score for the Pearson's correlation is computed, which is noted as $\overline{\rho}$. If $\gamma = \tau'$, the aggregated score for the Kendall's correlation is computed, noted as $\overline{\tau}'$. In Tables III, IV, and V, the objective measures are displayed in descending order according to $\overline{\rho}$.

The statistical significance of the difference between aggregated score couples is also tested in the Fisher-z domain [73]. Also for this statistical test, the two-tailed t value for $\alpha = 0.05$ is used as significance threshold. The smallest statistically significant differences are shown by columns A and B in Tables III, IV, and V. The aggregated score $\overline{\rho}$ for measure A (see symbol reported in column A) is significantly different to the aggregated score for measure B (same symbol in column B) and it is not significantly different to the measures listed in between. Taking as example Table III, the 2f-model (($\phi$) column A) differs significantly from SI-SA2f (($\phi$) column B) and from all other following measures to the end of the list.

## V. CORRELATION ANALYSIS: RESULTS

The following presentation of the results is divided into three parts considering: BAQ in the audio coding domain (Section V-A, Table III); BAQ in the source separation domain (Section V-B, Table IV); artifacts-only ratings in source separation (Section V-C, Table V). The aggregated scores $\overline{\rho}$ and $\overline{\tau}'$ for some selected measures are summarized in Table II.

### A. Results for BAQ in the Audio Coding Domain (Table III)

In the audio coding domain, the best aggregated scores are exhibited by the 2f-model ($\overline{\rho} = 0.90, \overline{\tau}' = 0.91$). SI-SA2f shows similar aggregated scores ($\overline{\rho} = 0.87, \overline{\tau}' = 0.89$), but with remarkable differences, especially for pre-echoes, as shown in Fig. 1. Considering the aggregated scores, the novel SI-SA2f outperforms the other artifacts-related measures: (SI-)SAR, APS, and LKR-PI. Both the 2f-model and SI-SA2f were designed in the source separation domain, but the underlying MOVs were developed in the audio coding domain.

TABLE III
BAQ: AUDIO CODING DOMAIN. EACH TABLE CELL REPORTS PEARSON'S $\rho$ AND MAPPED KENDALL'S $\tau'$, GIVEN IN PERCENT RELATIVE TO 1, E.G. 94 STANDS FOR 0.94. AGGREGATED SCORES $\overline{\rho}, \overline{\tau}'$ OVER ALL DATASETS IN THE TABLE ARE ALSO REPORTED. (†) VALUES ARE NOT USED FOR CALCULATING THE AGGREGATED SCORES (DATASET WAS USED IN THE DEVELOPMENT OF THE MEASURE). COLUMNS A AND B REPORT STATISTICALLY SIGNIFICANT DIFFERENCES, E.G., 2F-MODEL (($\phi$) COLUMN A) DIFFERS SIGNIFICANTLY FROM SI-SA2F (($\phi$) COLUMN B)

| | | Birdies | | Tonality M. | | BW Lim | | Unmasked N. | | Pre-echoes | | USAC t1 | | USAC t3 | | Aggregated scores | | A | B |
|---|---|---|---|---|---|---|---|---|---|---|---|---|---|---|---|---|---|---|---|
| 1 | 2f-model | 94* | 95* | 65* | 71* | 98* | 98* | 71* | 75* | 98* | 98* | 84* | 87* | 82* | 88* | 90 | 91 | $\phi$ | |
| 2 | SI-SA2f | 94* | 95* | 68* | 68* | 98* | 98* | 71* | 73* | 89* | 93* | 84* | 87* | 80* | 85* | 87 | 89 | ⋈ | $\phi$ |
| 3 | ADB | 92* | 92* | 64* | 74* | 96* | 97* | 73* | 76* | 93* | 92* | 77* | 82* | 81* | 88* | 86 | 88 | • | |
| 4 | HAAQI | 88* | 90* | 68* | 62* | 80* | 80* | 76* | 72* | 92* | 91* | 86* | 90* | 83* | 88* | 84 | 84 | ∝ | ⋈ |
| 5 | NMR | 81* | 86* | 81* | 92* | 96* | 96* | 81* | 87* | 88* | 87* | 70* | 74* | 70* | 76* | 83 | 88 | ∝ | |
| 6 | WinModDiff1 | 80* | 92* | 41* | 46* | 91* | 98* | 49* | 51* | 97* | 95* | 88* | 88* | 76* | 81* | 82 | 87 | ▽ | • |
| 7 | AvgModDiff1 | 81* | 94* | 42* | 46* | 91* | 98* | 51* | 52* | 94* | 89* | 89* | 89* | 78* | 83* | 81 | 87 | ▽ | |
| 8 | PEAQ ODG | 46* | 47* | 84* | 85* | 95* | 97* | 73* | 74* | 89* | 86* | 79* | 89* | 72* | 77* | 81 | 84 | ▽ | |
| 9 | ViSQOLAudio | 70* | 73* | 75* | 70* | 79* | 82* | 69* | 66* | 88* | 83* | 83* | 89* | 75* | 80* | 78 | 77 | × | ∝ |
| 10 | PESQ | 74* | 79* | 77* | 68* | 69* | 93* | 78* | 74* | 83* | 94* | 81* | 81* | 57* | 85* | 75 | 84 | ⊖ | ▽ |
| 11 | APS | 95* | 98* | 87* | 82* | 84* | 95* | 59* | 65* | 51* | 64* | 38* | 40* | 66* | 75* | 75 | 82 | ⊛ | |
| 12 | PEMO-Q ODG | 93* | 93* | 62* | 58* | 82* | 97* | 49* | 56* | 68* | 79* | 63* | 75* | 55* | 72* | 72 | 82 | ∓ | × |
| 13 | OPS | 86* | 88* | 65* | 54* | 91* | 95* | 50* | 52* | 57* | 65* | 63* | 64* | 60* | 63* | 71 | 74 | ⊘ | |
| 14 | VISQOLAudioV3 | 37* | 45* | 72* | 80* | 89* | 91* | 64* | 55* | 61* | 63* | 72* | 78* | 71* | 78* | 69 | 73 | ⊠ | ⊖ |
| 15 | POLQAv1 | 79* | 86* | 74* | 68* | 78* | 80* | 68* | 69* | 28 | 33 | 75* | 74* | 66* | 74* | 69 | 72 | ⊠ | |
| 16 | fwSNRseg | 71* | 73* | 6 | 8 | 91* | 94* | 35* | 35* | 67* | 69* | 74* | 77* | 79* | 85* | 67 | 71 | ⊠ | ⊛ |
| 17 | LKR-PI | 95* | 95* (†) | 48* | 53* | 60* | 80* | 50* | 49* | 59* | 72* | 36* | 59* | 65* | 77* | 65 | 74 | ◁ | ∓ |
| 18 | IPS | 58* | 63* | 37* | 47* | 88* | 95* | 40* | 49* | 73* | 81* | 69* | 76* | 67* | 73* | 65 | 74 | ◁ | |
| 19 | SI-SDR | 46* | 54* | 66* | 63* | 82* | 92* | 56* | 61* | 68* | 76* | 44* | 47* | 70* | 74* | 63 | 70 | $\theta$ | ⊘ |
| 20 | SI-SAR | 46* | 54* | 66* | 63* | 82* | 92* | 56* | 61* | 68* | 76* | 44* | 47* | 70* | 74* | 63 | 70 | $\theta$ | |
| 21 | POLQAv3 | 76* | 84* | 57* | 65* | 73* | 91* | 61* | 53* | 21 | 37* | 61* | 64* | 54* | 68* | 60 | 70 | $\beta$ | ⊠ |
| 22 | SDR | 34* | 42* | 58* | 55* | 68* | 89* | 52* | 58* | 66* | 77* | 50* | 53* | 70* | 75* | 58 | 67 | $\beta$ | ◁ |
| 23 | SAR | 34* | 42* | 58* | 55* | 68* | 89* | 52* | 58* | 66* | 77* | 50* | 53* | 70* | 75* | 58 | 67 | $\beta$ | |
| 24 | POLQAv2 | 76* | 79* | 47* | 61* | 57* | 60* | 59* | 34 | 35* | 24 | 62* | 67* | 42* | 61* | 56 | 57 | $\gamma$ | $\theta$ |
| 25 | dLLR | 2 | 5 | 13 | 4 | 67* | 79* | 12 | 5 | 85* | 78* | 60* | 65* | 65* | 71* | 49 | 51 | $\gamma$ | $\beta$ |
| 26 | WEnets PESQ | 22 | 24 | 2 | 4 | 49* | 38* | 8 | 12 | 23 | 16 | 29* | 36* | 5 | 3 | 20 | 19 | $\Omega$ | $\gamma$ |
| 27 | SIR | 40* | 44* | 0 | 0 | 0 | 0 | 31* | 31 | 44* | 44* | 0 | 0 | 0 | 0 | 17 | 18 | $\Omega$ | |
| 28 | SI-SIR | 0 | 0 | 0 | 0 | 0 | 0 | 0 | 0 | 0 | 0 | 0 | 0 | 0 | 0 | 0 | 0 | | $\Omega$ |

The lowest correlation coefficients observed for the 2f-model are for the listening test on tonality mismatch ($\overline{\rho} = 0.65, \overline{\tau}' = 0.71$), for which also the underlying MOVs (ADB and WinModDiff1) show weak correlation. This is one of the most challenging listening tests to be predicted in this domain, with only NMR, PEAQ ODG, and APS showing $\overline{\rho}$ and $\overline{\tau}' > 0.80$. Even more challenging is the listening test on unmasked noise, for which only NMR shows $\overline{\rho}$ and $\overline{\tau}' > 0.80$.

The four considered PEAQ MOVs are among the top ten aggregated scores, showing higher aggregated scores than their combination inside the PEAQ ODG, as also observed in [21].

Among the top five aggregated scores, three are achieved by measures not explicitly calibrated for audio coding, i.e. 2f-model, SI-SA2f, and HAAQI. Ignoring the MOVs, the best measures from the audio coding domain are met on Rank 8 (PEAQ ODG) and 9 (ViSQOLAudio). The different versions of POLQA are on Rank 15 (v1), 21 (v3) and 24 (v2).

WEnets PESQ is not able to mimic PESQ in any test.

TORCOLI et al.: OBJECTIVE MEASURES OF PERCEPTUAL AUDIO QUALITY REVIEWED 1537TABLE IV
BAQ: SOURCE SEPARATION DOMAIN. EACH TABLE CELL REPORTS PEARSON'S $\rho$ AND MAPPED KENDALL'S $\tau'$, GIVEN IN PERCENT RELATIVE TO 1. (†) VALUES ARE NOT USED FOR CALCULATING THE AGGREGATED SCORES. COLUMNS A AND B REPORT STATISTICALLY SIGNIFICANT DIFFERENCES

| | | PEASS OPS LT | | PEASS BAQ | | SiSEC08 | | SASSEC | | SiSEC18 | | SAOC DB | | Aggregated scores | | A | B |
|---|---|---|---|---|---|---|---|---|---|---|---|---|---|---|---|---|---|
| 1 | 2f-model | 87* | 83* | 80* | 76* | 92* | 88* | 88* | 90* (†) | 83* | 82* | 84* | 79* | 86 | 82 | ★ | |
| 2 | ADB | 86* | 83* | 76* | 73* | 91* | 87* | 80* | 83* | 80* | 79* | 79* | 75* | 83 | 81 | ★ | |
| 3 | PEMO-Q ODG | 84* | 81* | 79* | 77* | 76* | 75* | 77* | 82* | 78* | 81* | 67* | 65* | 77 | 78 | ⋌ | ★ |
| 4 | PESQ | 82* | 69* | 73* | 53* | 83* | 75* | 70* | 83* | 73* | 74* | 74* | 58* | 76 | 70 | ⋌ | |
| 5 | POLQAv2 | 81* | 84* | 68* | 74* | 72* | 58* | 64* | 60* | 82* | 84* | 71* | 54* | 74 | 71 | ◇ | |
| 6 | POLQAv3 | 84* | 86* | 73* | 77* | 73* | 53* | 63* | 57* | 78* | 79* | 68* | 50* | 74 | 70 | ◇ | |
| 7 | POLQAv1 | 84* | 81* | 70* | 71* | 75* | 56* | 61* | 55* | 77* | 78* | 72* | 58* | 74 | 68 | ◇ | |
| 8 | fwSNRseg | 80* | 77* | 71* | 63* | 83* | 84* | 17 | 45* | 80* | 81* | 62* | 55* | 69 | 70 | ⊘ | ⋌ |
| 9 | AvgModDiff1 | 58* | 58* | 68* | 69* | 76* | 80* | 72* | 92* | 71* | 79* | 70* | 76* | 69 | 78 | ⊘ | |
| 10 | HAAQI | 72* | 59* | 62* | 38 | 82* | 74* | 65* | 71* | 59* | 65* | 71* | 52* | 69 | 61 | ⊘ | |
| 11 | OPS | 92* | 92* (†) | 86* | 89* | 65* | 43* | 59* | 58* | 57* | 47* | 54* | 47* | 67 | 63 | ⊘ | ◇ |
| 12 | dLLR | 55* | 67* | 54* | 60* | 67* | 75* | 85* | 85* | 70* | 72* | 62* | 66* | 67 | 72 | ⊘ | |
| 13 | ViSQOLAudio | 64* | 60* | 58* | 54* | 73* | 69* | 68* | 68* | 71* | 69* | 66* | 68* | 67 | 65 | ⊘ | |
| 14 | WinModDiff1 | 62* | 54* | 62* | 58* | 65* | 62* | 74* | 92* | 65* | 73* | 68* | 73* | 66 | 72 | α | |
| 15 | NMR | 61* | 62* | 36 | 36 | 70* | 62* | 83* | 84* | 62* | 67* | 67* | 63* | 65 | 65 | α | |
| 16 | SI-SA2f | 16 | 17 | 26 | 32 | 91* | 83* | 82* | 83* (†) | 75* | 78* | 70* | 66* | 65 | 62 | α | |
| 17 | PEAQ ODG | 66* | 72* | 60* | 55* | 75* | 68* | 71* | 83* | 55* | 61* | 49* | 51* | 63 | 66 | ◁ | |
| 18 | SDR | 61* | 67* | 48* | 50* | 63* | 66* | 69* | 74* | 69* | 75* | 67* | 71* | 63 | 68 | ◁ | |
| 19 | SIR | 70* | 76* | 54* | 59* | 43* | 47* | 65* | 68* | 61* | 62* | 56* | 60* | 59 | 63 | θ | ⊘ |
| 20 | SI-SDR | 33* | 54* | 26 | 43* | 63* | 68* | 68* | 74* | 66* | 75* | 67* | 71* | 56 | 65 | ω | α |
| 21 | SI-SIR | 61* | 66* | 52* | 53* | 41* | 46* | 67* | 68* | 56* | 58* | 54* | 60* | 56 | 59 | ω | |
| 22 | VISQOLAudioV3 | 38* | 33 | 33 | 29 | 59* | 72* | 66* | 74* | 66* | 68* | 43* | 49* | 52 | 57 | Ω | ◁ |
| 23 | APS | 13 | 5 | 33 | 25 | 73* | 76* | 69* | 73* | 56* | 59* | 48* | 46* | 51 | 51 | Ω | |
| 24 | SI-SAR | 14 | 21 | 9 | 12 | 62* | 66* | 65* | 73* | 65* | 73* | 65* | 69* | 50 | 56 | Ω | θ |
| 25 | LKR-PI | 54* | 62* | 23 | 3 | 58* | 58* | 48* | 59* | 48* | 41* | 55* | 54* | 48 | 48 | Ω | |
| 26 | SAR | 1 | 11 | 2 | 4 | 62* | 66* | 61* | 68* | 66* | 71* | 67* | 71* | 47 | 53 | Ω | |
| 27 | IPS | 77* | 72* (†) | 59* | 59* | 41* | 31* | 32* | 29* | 11 | 4 | 28* | 30* | 45 | 40 | Ω | ω |
| 28 | WEnets PESQ | 9 | 18 | 9 | 18 | 30* | 16 | 61* | 62* | 9 | 16 | 23* | 10 | 25 | 25 | | Ω |

Finally, non-perceptual, signal-based measures dominate the lower third of the ranking: (SI-)SDR and (SI-)SAR.

SI-SIR shows no correlation, as expected, since there is no interference present in this domain. On the other hand, IPS strongly correlates with some of the artifact types, such as BW Lim and Pre-echoes. This can be observed in detail in Fig. 2 and is surprising as no interfering signal is present in this case. As expected, $q^{\text{interf}}$ is always maximum ($= 1.0$) for these signals. PEASS Version 2 takes $q^{\text{interf}}$, $q^{\text{artif}}$, and $q^{\text{global}}$ as inputs for the final 2-layers neural network producing the IPS. Here, $q^{\text{global}}$ seems to have a decisive impact on the final IPS, even if IPS should be only related to the interference.

Over all measures, best aggregated scores are achieved on the BW Lim dataset, while the worst ones are achieved on Tonality Mismatch and Unmasked Noise.

### B. Results for BAQ in the Source Separation Domain (Table IV)

In the source separation domain, the best measure is again the 2f-model ($\overline{\rho} = 0.86$ and $\overline{\tau}' = 0.82$), even if with slightly lower aggregated scores than in the audio coding domain. The 2f-model shows similar performance to ADB, with which no significant difference is observed. The other aggregated scores up and $\overline{\tau}' = 0.62$), showing particularly low correlation with the listening tests PEASS OPS LT and PEASS BAQ. For these listening tests, (SI-)SIR show some significant correlation (up to $\overline{\rho} = 0.70$ and $\overline{\tau}' = 0.76$). This suggests that the different levels of the interferer strongly contributed to BAQ in these listening tests.

TABLE V
ARTIFACTS ONLY: SOURCE SEPARATION DOMAIN. TASK:" RATE THE QUALITY IN TERMS OF ABSENCE OF ADDITIONAL ARTIFICIAL NOISE IN EACH TEST SIGNAL." EACH TABLE CELL REPORTS PEARSON'S $\rho$ AND MAPPED KENDALL'S $\tau'$, GIVEN IN PERCENT RELATIVE TO 1

|    |              | PEASS APS LT | A | B |
|----|--------------|--------------|---|---|
| 1  | APS          | 89* 91* (†)  | ★ |   |
| 2  | LKR-PI       | 77* 80* (†)  | ● |   |
| 3  | SI-SA2f      | 64* 67*      | ◇ | ★ |
| 4  | AvgModDiff1  | 55* 46*      | λ |   |
| 5  | SAR          | 48* 55*      | ω |   |
| 6  | HAAQI        | 39* 18       |   | ● |
| 7  | WinModDiff1  | 38* 38*      |   |   |
| 8  | PEAQ ODG     | 37* 6        |   |   |
| 9  | IPS          | 33* 43*      |   |   |
| 10 | NMR          | 32* 39*      |   |   |
| 11 | 2f-model     | 27 12        |   | ◇ |
| 12 | SI-SIR       | 27 32        |   |   |
| 13 | SIR          | 26 31        |   |   |
| 14 | PESQ         | 26 4         |   |   |
| 15 | POLQAv1      | 23 16        |   |   |
| 16 | WEnets PESQ  | 23 25        |   |   |
| 17 | POLQAv3      | 21 7         |   |   |
| 18 | PEMO-Q ODG   | 19 16        |   |   |
| 19 | SI-SAR       | 19 20        |   |   |
| 20 | fwSNRseg     | 18 4         |   |   |
| 21 | VISQOLAudioV3| 17 21        |   |   |
| 22 | SDR          | 16 15        |   |   |
| 23 | POLQAv2      | 15 10        |   | λ |
| 24 | ADB          | 13 13        |   |   |
| 25 | OPS          | 11 3         |   |   |
| 26 | dLLR         | 10 9         |   |   |
| 27 | SI-SDR       | 1 14         |   | ω |
| 28 | ViSQOLAudio  | 0 16         |   |   |

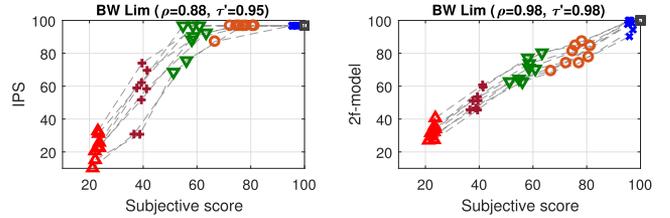

Fig. 2. Interference-related Perceptual Score (IPS, left) and 2f-model (right) predicting BAQ for the listening test BW Lim. No interference signals are present here, but IPS shows clear correlation with BAQ regardless.

The non-perceptual, signal-based measuring methods, such as BSSEval, dominate again the lower third of the ranking. WEnets PESQ is able to mimic PESQ only for SASSEC and not for all other listening tests.

### C. Results for the Artifacts-Only Scores for the Source Separation Domain (Table V)

In source separation, it is often of interest to assess the interferer reduction and the presence of distortions, artifacts, and colorations independently rather than jointly [55], [75]. This can give useful diagnostic information, e.g., for supporting the interpretation of a listening test [76] or for controlling the amount of interferer reduction such that a desired artifacts-related quality level is met [77]. As far as perceptual scores assessing exclusively artifacts, only one dataset is available (PEASS APS LT), in which listeners rated the quality in terms of absence of additional artificial noise. Conclusions should be corroborated on more data for this scenario.

APS yields the highest scores, but it was trained on this dataset. LKR-PI shows no statistically significant difference with APS, but the dataset was used as validation set for this measure. SI-SA2f is the first system in the list for which the data was completely unknown ($\rho = 0.64$ and $\tau' = 0.67$). Furthermore, the novel SI-SA2f outperforms the remaining artifacts-related measures (SI-)SAR. Also, SI-SA2f was shown to consistently outperform APS and the other artifacts-related measures in the other considered cases (Table III and Table IV). AvgModDiff1 follows with $\rho = 0.55$ and $\tau' = 0.46$. This MOV assesses the differences in the temporal modulation of the loudness envelopes of the reference and the test signal. These differences can be an indicator of the presence of artifacts-related distortions.

All other considered measures either generally assess the differences between test and reference signal or are tailored to assess other specific aspects (e.g. interferer). As expected, these measures perform poorly on this dataset.

to Rank 10 ($\overline{\rho} \geq 0.69; \overline{\tau}' \geq 0.61$) are achieved by measures not from the source separation domain. Notably, POLQA performs better in this domain than in audio coding.

It might seem counterintuitive at first, but it is to be expected that measures designed to assess only artifacts (SI-SA2f, APS, (SI-)SAR, and LKR-PI) perform better in the audio-coding domain even if designed for source separation. In both domains, listeners assessed BAQ and judged any and all perceived differences between the reference and the test signals, but only the signals in the source separation domain present also non-artifact related differences with the reference, i.e. interferer of varying level. In other words, assessing only artifacts and assessing BAQ are the same task in the considered listening tests about audio coding, while they are very different tasks in source separation. The best measure of these is again SI-SA2f ($\overline{\rho} = 0.65$

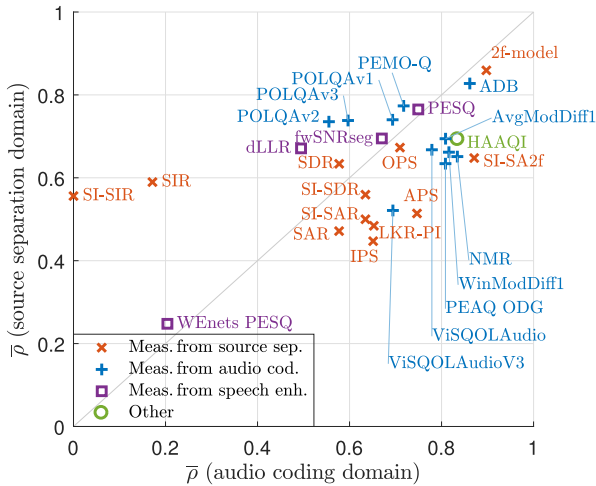

Fig. 3. Aggregated Pearson's $\overline{\rho}$ between the considered objective measures and the ground-truth BAQ scores for the source separation domain and the audio coding domain. Domain-independent models would lie on the diagonal, possibly best in the top right corner, i.e. approaching $\overline{\rho} = 1$ for both domains.

## VI. DISCUSSION

Fig. 3 gives an overview of the aggregated Pearson's correlation scores observed for BAQ in the source separation domain against the audio coding domain. It can be observed that the 2f-model is the best measure in both domains, showing a high degree of domain independence. The 2 MOVs used by the 2f-model perform well also on their own (especially ADB), but their combination in the 2f-model outperforms their individual performance. Also PESQ and PEMO-Q ODG show medium-to-high correlation scores on both domains ($0.72 \leq \overline{\rho} \leq 0.77$), hence being possible second best choices. In fact, 2f-model, ADB, PESQ, and PEMO-Q populate the top right corner of Fig. 3. These results indicate the possibility of an application domain independent model. For this purpose, the importance of two components is highlighted: 1) an accurate perceptual model (PEAQ MOVs, PESQ, and PEMO-Q are almost all older than 20 years and still show their validity in most cases) and 2) varied training data comprising a large variety of different audio material with different distortion types.

More work is still to be done on both fronts. Purely signal-based measurement methods without perceptual aspects, e.g. SDR, perform worst in both domains, but perceptually-motivated measures also show their limits, especially with more modern distortion types, e.g. tonality mismatch and unmasked noise (suboptimal parametric coding of the higher frequency bands). On the other hand, more research and training data are needed for improving the correlation performance, especially in the source separation domain, where the correlation scores are generally lower. OPS from the PEASS toolkit shows on the source separation datasets a high correlation only on the PEASS OPS LT and PEASS BAQ dataset. This may indicate that the training dataset was too limited or over-fitted. Moreover, PEMO-Q is used internally by OPS as perceptual model, but PEMO-Q performs overall significantly better than OPS in the source separation domain.

Besides 2f-model, ADB, PESQ, and PEMO-Q, the performance of the remaining measures show a certain domain dependence or medium to low correlation in general. Some (e.g. PEAQ ODG) reveal lower performance on the *unknown* domain. Surprisingly, others achieve better performance on the *unknown* domain (e.g. POLQA).

When assessing exclusively the perceptual relevance of artifacts, SI-SA2f seems to be the most promising measure. Compared to the measures performing best in this task (APS and LKR-PI, Table V), SI-SA2f was not calibrated or validated on this dataset and showed better robustness on the other datasets (Table III and IV). More test data is needed, however, for this special case.

Similar but not identical results were observed for the 2f-model and SI-SA2f in the audio coding experiments. Only artifacts-related degradations are present in the listening tests for this domain. Intuition would suggest that the results should be identical for the 2f-model and SI-SA2f. However, the two measures rely on different definitions of artifacts. The 2f-model leverages a perceptually motivated definition as per PEAQ MOVs, which is a well established approach in the audio coding community. SI-SA2f uses the purely signal-based definition of BSSEval, which is a popular approach in the source separation community. Here, the artifacts are defined as the projection error when trying to explain the separated target source signal by a linear projection of the original source signals onto the signal mixture. Everything that cannot be explained by a filtered version of the original source signals is considered as artifact. This is not necessarily congruent with the perceptual definition of artifacts and highlights the importance of a more perceptually-motivated signal decomposition.

Finally, WEnets PESQ showed very low correlation in almost all experiments. It has to be noted that WEnets PESQ operates without reference signal, so the task for this measure is significantly more difficult than for all other measures. The original work reports Pearson correlation of 0.97 with PESQ [63], where training and testing signals were speech items processed by different speech codecs followed by noise suppression. This type of material fits PESQ original domain, but not many of our experiments, where, e.g. also music is present and very different processes. This domain mismatch brings much more dramatic consequences for the purely data-driven method (WEnets PESQ) with respect to the perceptually-motivated steps of PESQ.

## VII. CONCLUSION

Aggregating the correlation coefficients from 13 listening tests including a range of application domains and distortions, the possibility of a domain-independent model for predicting Basic Audio Quality (BAQ) has been shown. However, only a very limited number of the considered state-of-the-art tools exhibit domain independence along with high correlation scores. The source separation domain appears to be a particularly challenging application domain, with only two measures showing aggregated scores $\geq 0.80$. For this application domain both the interferer level and artifacts and colorations contribute to the final BAQ.

The 2f-model showed the best aggregated correlation for both the audio coding domain ($\overline{\rho} = 0.90$, $\overline{\tau}' = 0.91$) and the source separation domain ($\overline{\rho} = 0.86$, $\overline{\tau}' = 0.82$). This model uses perceptual features from the audio coding domain developed more than 20 years ago as part of PEAQ. Their combination is trained with data from the source separation domain. This mix of domains, the validity of the underlying perceptual model, and the varied training data appear to be successful strategies in addressing a big range of audio quality degradations. The main output from PEAQ itself shows good correlation only for the domain on which it was trained and it is also often outperformed by the individual underlying features. This suggests that the auditory models themselves still hold their validity (with the only exception of parametric bandwidth extension), while more heterogeneous data is needed for training and calibration. The same conclusion can be drawn when analyzing OPS, which performs well only on the signals on which it was trained.

Besides the 2f-model, also ADB, PESQ, and PEMO-Q show similar correlation scores on both domains along with medium-to-high correlation ($0.72 \leq \overline{\rho} \leq 0.77$), as it can also be observed in Fig. 3.

DNN-based methods are still in the early phase of development, mostly because of the difficulty in collecting large amount of ground-truth subjective scores. Training DNNs to predict the output of an existing measure (e.g. so to make it non-intrusive) is an alternative that still needs to be proven robust in the actual application.

If only artifacts and colorations are to be assessed, regardless of the interferer level, the most promising measure seems to be SI-SA2f, which is a novel measure based on the 2f-model and preceded by a BSSEval-based signal decomposition.

TORCOLI *et al.*: OBJECTIVE MEASURES OF PERCEPTUAL AUDIO QUALITY REVIEWED  1541